\documentstyle[dina4,12pt]{article}

\begin{document}
\bibliographystyle{unsrt}
\textwidth 800pt
\large
\begin{center}
\underline{Slow relaxation and phase space properties
} \\
\underline{ of a conservative system with many degrees of freedom
}
\vspace{2cm}\\ \large S. \vspace{0.5cm}Flach\\
\normalsize
Department of Physics, Boston University, 590 Commonwealth Avenue,\\
Boston, Massachusetts 02215, \small flach@buphy.bu.edu \large
\vspace{1cm}
\\
\large G. Mutschke \vspace{0.5cm} \\
\normalsize
Research Center Rossendorf, P.O. Box 51 01 19, 01314 Dresden, \\
Federal Republic
of Germany
 \vspace{1cm}
\\
\end{center}
\normalsize
ABSTRACT \\
The object of our study is the one-dimensional discrete $\Phi^4$ model.
We compare two equilibrium properties by use of
molecular dynamics simulations: the
Lyapunov spectrum and the time dependence of displacement-displacement
and energy-energy correlation functions.
Both properties imply the existence of a dynamical crossover of the
system at the same temperature. This correlation holds for two
rather different regimes of the system - the displacive and intermediate
coupling regimes. These results imply a deep connection between
slowing down of relaxations and phase space properties of complex systems.
\vspace{0.5cm}
\newline
PACS number(s): 03.20.+i ; 63.20.Pw ; 63.20.Ry
\newline
{\sl Physical Review E}, in press.
\newline
Date: 03/14/94
\newpage

\section{Introduction}

The classical statistical mechanics of macroscopic systems in
equilibrium essentially
uses the ergodicity conjecture, i.e. time averages are replaced
by phase space averages. The nongeneric occurence of integrable
systems and the existence of stochastic webs in phase space of
nonintegrable systems with $N \geq 3$ ($N$ is the number of degrees
of freedom) \cite{eaj90} together with Boltzmann's approach
(cf. e.g. \cite{jll93}) provides an intuitive
explanation of the ergodicity conjecture.
However the dynamics of nonlinear macroscopic systems shows up with
rather complex properties so that further details of nonlinear dynamics
have to be exploited. In this contribution we deal especially with
properties of slow relaxations. Common examples could be critical
slowing down near second order phase transitions \cite{sm76}, freezing near
the liquid-glass transition \cite{wg91}. In these problems one has to deal
with dynamics on different time scales. The success of phenomenological
and semiphenomenological theories to describe slow relaxations
in those systems does not alter the fact that we are far from completely
understanding the underlying microscopic dynamics.

Modern theory of nonlinear dynamics provides us with several
useful results. First we mention the KAM theorem \cite{eaj90}. It states
that if an integrable system is slightly perturbed with a nonintegrable
perturbation, there exists a set consisting of $N$-dimensional
tori close to the tori of the unperturbed integrable system. The
set in the perturbed system is nowhere dense but forms a large
part of the phase space (i.e. the measure of the complement of the set
tends to zero as the perturbation is lifted). If the perturbation
strength overcomes a finite value, most of the perturbed tori are
destroyed. The KAM theorem deals only with the possibility
of a nonintegrable system to evolve on regular trajectories (tori).
In that sense KAM makes no statements about finite time stabilities.
Second we mention the Nekhoroshev theorems \cite{nnn71}.
These theorems deal with
finite time stabilities. They provide us with lower bounds on time
scales on which the nonintegrable system evolves on a trajectory
close to a regular one. Finally we mention the numerical evidence
for the existence of strong stochasticity thresholds SST (in the
strength of the perturbation) \cite{pl90}.
Below the SST the system's trajectory
evolves mainly along resonances in phase space. Above the SST
the trajectory evolves across resonances thus speeding up
the relaxation of the system, which can be roughly brought into
connection with the time the system's trajectory needs to
cover a major part of the available phase space.

A subtle point in the application of the above results to macroscopic
systems is the dependence of different threshold values on
the number of degrees of freedom. Despite controversal opinions
there seems to be some agreement that neither the KAM-tori nor
the Nekhoroshev finite time regularity survive in the limit
$N \rightarrow \infty$ \cite{mp93}. In other words, those properties are
supressed to regions of almost zero energies per degree of freedom
(temperature) in the thermodynamic limit. Only the SST seems to survive.
The increase of the energy per degree of freedom
in a nonlinear nonintegrable system is equivalent to the increase
of the strength of a certain nonintegrable perturbation. Then it could
be possible that at certain finite energies per degree of freedom the
system will be close to another integrable system. For instance it is
possible for certain systems to increase the energy per degree of freedom
to infinity and become infinitely close to an integrable system.
Thus we would not a priori rule out the applicability of the KAM and
Nekhoroshev results to macroscopic systems at finite temperature.

In this work we present results of numerical experiments for
a simple one-dimensional lattice model. We show that both the
relaxation times and Kolmogorov-Sinai entropy (KSE) are sensitive
to the existence of a dynamical 'phase transition'. In other
words we can deduce changes in relaxational properties by measuring
the KSE and changes in the type of phase space trajectories by
measuring relaxation times. The interpretation of our results is
(in our opinion) closely connected to the thresholds discussed
in the above paragraph. It is also useful in detecting new kinds
of elementary excitations (quasiparticles) in complex systems,
as we will demonstrate by analysing two particular model realizations.

\section{Model, numerical methods}

We study a  d=1 dimensional discrete classical model
given by the Hamiltonian
\begin{equation}
H= \sum_{l=1}^N \left[ \frac{1}{2}P_l^2 + \frac{1}{2}C(X_l - X_{l-1})^2
+ V(X_l) \right] \;\;\;. \label{1}
\end{equation}
$P_l$ and $X_l$ are canonically conjugated momentum and displacement
of the $l-$th particle, where $l$ marks the number of the unit cell.
$C$ measures the interaction to the nearest neighbour particles.
All variables are dimensionless. The mass of the particles is
equal to unity. $N$ is the total number of particles.
The nonlinearity appears in the 'on-site'
potential $V(x)$ which is of the $\Phi^4$ type:
\begin{equation}
V(x) = V_{\Phi^4}(x) = \frac{1}{4}(x^2 - 1)^2 \;\;\;\;,\hspace{2cm} \label{2}
\end{equation}
The barrier height of this double well potential is $\Delta = 0.25$.
The interaction parameter $C$ and the energy per particle $E/N$
($E$ is the total energy of the system)
are the two parameters of the system. The temperature $T$ (mean squared
velocity) is then given through a virial theorem.
The classification of the system behaviour in the $(C,T)$ plane
turns out to be rather complex. First we mention the existence
of a critical point (second order phase transition) on the line
$(C,T_c=0)$ \cite{ss77}.
The order parameter is $<X>=\sum_l X_l$.
 For finite temperatures below the
value of $\Delta$ one finds: Ising-like (order-disorder)
behaviour for $C \ll 1$; displacive (continuous) behaviour for
$C \geq 1$; and a subtle intermediate behaviour in between the
two previous $C$ ranges. For temperatures tending to infinity
the system behaves like uncoupled quartic oscillators with canonical
energy distribution. For any finite value of $C$ the correlation length $\xi$
will decrease from infinity ($T=0$) to zero ($T=\infty$). Fixing
the temperature one finds zero correlation length for $C=0$.
Increasing $C$ leads to an increase of $\xi$. For $C \rightarrow \infty$
the correlation length tends to infinity. This is due to the fact
that the critical region around $T_c=0$ increases as the interaction
is increased.

The slow relaxational dynamics of our system is in part our focus.
The simplest way
to describe it would be to consider the
correlation function
$S_{A_lA_k}(\omega)$
\begin{eqnarray}
S_{A_lA_k}(t) = \langle A_l(t)\cdot A_k\rangle \;\;,\;\;
S_{A_lA_k}=\langle A_l\cdot A_k\rangle \;\;, \label{2-2} \\
S_{A_lA_k}(\omega) = S_{A_lA_k}(z=\omega + i0) \;\;, \nonumber \\
F(z)=LT[F(t)]=\frac{1}{i}\int_0^{\infty}{\rm d}t e^{izt}\cdot F(t) \;\;.
\nonumber
\end{eqnarray}
Here $\langle ... \rangle$ denotes standard canonical
average and LT[...] means Laplace transformation. The local microscopic
variable $A_l$ could be any combination of the canonical
variables setting up our desired Hamiltonian in \ref{1},\ref{2}.
The imaginary part of the susceptibility is then defined as
\begin{equation}
\chi_{A_lA_k}^{''}(\omega)=\frac{1}{2}\omega
\cdot S_{A_lA_k}(\omega)/S_{A_lA_k}(t=0) \label{2-3}
\end{equation}
and can be studied on a logarithmic frequency scale,
as commonly done to study slow relaxations
in glass dynamics \cite{wg91}.
Since the order parameter at the phase transition at $T_c=0$
is $<X>=\sum_l X_l$, we expect to observe critical slowing down in both
$S_{X_lX_k}$ and $\chi_{X_lX_k}^{''}$. The half width of a central peak around
$\omega = 0$ in $S_{X_lX_k}(\omega)$ or the position of the corresponding low
frequency peak in $\chi_{X_lX_k}^{''}(\omega)$ could serve as an
inverse relaxation
time.
%
%

Another powerful and mathematically well defined method in studying
nonlinear dynamics of complex systems is the Lyapunov spectrum, meaning
in our case the set of $2N$ Lyapunov exponents of a one-dimensional
N-particle system, ordered with respect to their magnitude
$\{\lambda_1 \ge \lambda_2 \ge ... \ge \lambda_{2N} \}$.
For Hamiltonian systems that are considered here the spectrum is symmetric with
respect to zero, i.e. $\lambda_i = - \lambda_{2N-i+1}$ because of the
symplectic evolution in the tangent space \cite{via89}.
In the thermodynamic limit $N\to\infty$ the existence of a smooth distribution
of Lyapunov exponents has been numerically verified \cite{lprv87}.
Intensive quantities like a Lyapunov density
\begin{equation}
\lambda(x) = \lambda_{i/N}, \quad \frac{i}{N}\to x, \quad N\to\infty
\end{equation}
or the Kolmogorov-Sinai entropy per particle $S_{KS}$
\begin{equation}
 S_{KS} = \int\limits_0^1 \lambda(x)\; dx
\end{equation}
can be introduced
as useful quantities for the description of the system behavior.
In simplest cases it can be explicitely shown
that the value of the positive Lyapunov exponent, which characterizes
the correlation decay on short time scales (local instability), is also
connected (sometimes linearly) to a relaxation time of the system
\cite{gmz85}.
In other words the Lyapunov exponents determine the relaxational
behaviour of correlators. However in systems with many degrees of
freedom the connection between the Lyapunov exponents and the
time dependence of correlators on long time scales is not known.
In highly chaotic regimes the shape of the function $\lambda(x)$ is
always linear and independent of the special form of the Hamiltonian,
as it was proved by random matrix approximations \cite{pv86}.
As the temperature decreases, decreasing of the KSE occurs due to
an increase in the curvature of $\lambda(x)$ and a decrease of
$\lambda_1$ towards a less chaotic behavior.
In the low temperature regime the shape of $\lambda(x)$  around $x=0.5$
is nestled against the x-axis yielding a growing number of very small Lyapunov
exponents preparing the smooth transition to the
integrable case $S_{KS}=0$ at $T=0$.
{}From the special shape of the KSE as function
of the temperature we expect to get additional information about the longest
time scales of the system with respect to the shortest ones, i.e. about the
relaxation behavior.
A similar method was used in \cite{bc87} to detect a phase transition in a
2d-Heisenberg model and in \cite{pl90},\cite{mp93} to detect the SST.
%
%
To evaluate the properties of interest we used molecular dynamics methods.
The detailed explanation is given in \cite{fs93} for the time dependence
of correlators and in \cite{mb93} for the Lyapunov spectra.

\section{Results}

\subsection{$C=4$}

For $C=4$ we find the following scenario. With decreasing temperature
the inverse correlation length $1/\xi$ decreases. At $T \approx 0.35$
$\xi$ is of the order 200 \cite{fsss91},\cite{fs93}.
At that temperature a drastic change in
the temperature dependence of $1/\xi$ (on a linear scale)
takes place \cite{fs93}.
On a linear scale (Fig.1 in \cite{fs93}) it looks like $1/\xi$ becomes
zero below the crossover temperature. However that is not the case
(there is no phase transition at finite temperatures in those system).
Instead the correlation length stays finite at lower temperatures,
but it becomes very large.
A discontinuity
is seen in the inverse static susceptibility at this temperature.
The temperature dependence of the position $\omega_{\alpha'}$
of the low frequency relaxational peak
in $\chi_{X_lX_l}''(\omega)$ (cf \cite{fs93}) is shown in Fig.1.
Clearly a crossover behaviour
at $T \approx 0.3$ is observed.

In Fig. 2 we show the temperature dependence of the KSE. Again we
find a crossover behaviour around the same temperature $0.3$.
The KSE tends to zero by lowering the temperature to the crossover
value. Below the crossover the temperature dependence of the KSE is
seemingly drastically changed.

The interpretation of the excitation spectrum of the system
goes as following. At temperatures around $0.5$ and below kink-induced
relaxations become well separated from (still anharmonic) phonon
excitations \cite{fs92}.
The decrease of temperature leads to a decrease of the density of the kinks,
thus allowing us to detect their presence in the low frequency part of the
spectrum. Despite the discreteness of the system (lattice)
the kinks don't feel the negligible Peierls-Nabarro potential
(Peierls-Nabarro barrier is $\approx 4\cdot 10^{-8}$) and essentially
are moving as in the corresponding continuum system \cite{fw1}.
Because the density of the kink subsystem is low the collisions between
kinks become rare.
The increasing relaxation times appear because of lowering the
kink density. Only the motion of kinks can provide the system with
an equilibration channel. Let us test the applicability of the
phonon-kink picture where phonons and kinks are assumed to be
noninteracting with each other. Then it follows that the inverse correlation
length (or kink density) is proportional to the square root of
the inverse temperature multiplied with an exponent of $-E_k/T$, where
$E_k=2/3\sqrt{2C}$ is the minimum kink energy \cite{ckbt80}.
In Fig.3 we clearly see this law
to be realized. A rough estimation of the kink energy from the
slope in Fig.3 even yields the continuum value within 5\% \cite{fw1}.
This result also indicates that the above discussed dynamical crossover
is not detected in the simple temperature dependence of the relaxation
times and correlation length, but rather in more subtle dynamical
scaling properties \cite{fs93}. Consequently we expect the same to apply
to the KSE. We show in Fig.4 the realization of the same temperature
law as in Fig.3 for the correlation length. Thus the essential result
we find is that the rapid decrease of the KSE below $T=0.5$ indicates
the system to be close to an integrable one. The analysis of the
high frequency excitation spectrum as well as the low frequency relaxation
spectrum leads to the conclusion that the system can be described
by a mixture of nearly noninteracting phonons and kinks with corresponding
temperature dependent kink density.

\subsection{$C=0.1$}

For $C=0.1$ the scenario is changed. The correlation length increases
very slowly with decreasing temperature. At $T=0.1$ it is still of
the order of 5 lattice spacings \cite{fsss91},\cite{fs93}. The relaxation time
increases much more rapidly. In \cite{fsss91} it is seen that an analogous
crossover temperature as in the $C=4$ case seems to be reached at temperatures
around $T=0.1$. In Fig.5 we show the temperature dependence of
$\omega_{\alpha'}$ (cf. \cite{fs93}) as a function of temperature.
Indeed a strong slowing
down is observed around the above cited temperature.

In Fig.6 the KSE(T)-dependence is shown.
First we find a maximum in the KSE($T$)-dependence around
$T=0.5$ (for a discussion see \cite{mb93}).
Below that temperature the KSE rapidly decreases
with decreasing temperature.That indicates that for $T < 0.5$
one can again try to find an integrable system which is close
to the studied one. Secondly there is a step-like decrease in the KSE
at $T \approx 0.03$ (insert in Fig.6).

The interpretation of the excitation spectrum is not wellknown in that
case.
Let us start with the still present kink subsystem.
The change of the interaction
parameter $C$ from 4 to 0.1 is mainly affecting the movability
of kinks. That should happen because the Peierls-Nabarro Potential
the kink feels during its motion through the lattice has a barrier
height of approximately 0.164 \cite{fw1}. Thus the kinks are trapped by
the discreteness of the lattice. The radiation of energy by moving
kinks is also strong compared to the $C=4$ case. Then the
lattice site change of a kink becomes a hopping process without
strong correlations to previous site changes. Consequently
for same kink densities as in the $C=4$ case
a longer relaxation time of the displacement-displacement correlator
can be expected, as found in the numerical simulations \cite{fs93}.

Still the kink density decreases with temperature, so that more and
more different degrees of freedom are excited when lowering the
temperature. In contrast to the $C=4$ case the high-frequency part
of the displacement-displacement spectrum is far from being described
by (weakly interacting) phonons \cite{fs92},\cite{fs93}.
The spectrum in this frequency range is qualitatively very similar
to spectra of uncorrelated particles ($C=0$) \cite{yo70}.
However this seems to
be strange since one can estimate the interaction contribution for
$C=0.1$ and find that in the given temperature (energy) ranges
the coupling energy is comparable with the total energy \cite{fw2}.
To understand the nature of
this high-frequency part we show in Fig.7 the time dependence
of the local energy-energy correlator for different temperatures,
i.e. for $A_l=e_l=P_l^2/2+V(X_l)+ 0.25 C ((X_l-X_{l-1})^2+(X_{l+1}-X_l)^2)$.
If the kink excitations are the only localized ones, then we expect
a plateau to appear in the correlator. We could estimate the
height $h_{ee}$ of the plateau by knowing the kink energy $E_K\approx
0.258$ \cite{fw1} and the
kink density $1/\xi$: $h_{ee}=E_K^2/(2\xi)$. For all temperatures in
Fig.7 we find that we underestimate the height of the plateau by
30\%-50\%. Thus we have to conclude that other degrees of freedom in the
system are excited, which provide energy localization. The explanation
of the puzzle is very likely the existence of nonlinear localized
excitations (NLE) \cite{fw3},\cite{fw2}. These NLEs can be excited without the
presence of topologically induced kinks as well as in combination
with kinks.
The NLEs are (nearly) regular solutions of the nonlinear
translationally invariant lattice. A single NLE is described by a finite
set of fundamental frequencies and can be viewed as the excitation
of a finite set of nonlinear localized degrees of freedom. Thus
a given lattice which shows up with NLEs at finite temperatures
can be viewed as evolving (close to) on high-dimensional tori
in phase space for finite times.
The typical NLE for the case under study consists
out of three excited particles, one central (large amplitude)
and two neighbours (small amplitudes). Indeed the NLEs can be
observed in hypsometric plots in \cite{fs93}. In Figs.8,9 we show
two hypsometric plots which demonstrate the presence of NLEs
at the temperature $T=0.1$. Finally in Fig.10 we show $\chi_{e_le_l}^{''}$
versus frequency. The huge halfwidth of the low-frequency
relaxational peak (nearly 3 decades) indicates that several
relaxational processes are present - e.g. the kink hopping and
the NLE relaxation.
There are two intriguing facts which support the above given
interpretation of the spectrum. First it is known that the
NLEs (excluding the NLEs excited on kinks) have an existence
energy threshold \cite{fw3}. For $C=0.1$ this threshold has a value of about
$E=0.1$ \cite{fw3}. Since three particles are involved in the NLE,
it yields an energy of 0.03 per particle. This value comes rather
close to the above described step in the KSE at $T=0.03$. The
reason why the KSE $increases$ steplike if one heats the system
above the step temperature might be that below $T=0.03$ essentially
no NLEs are excited, so the system excites small amplitude phonons
which can have longer lifetimes compared to the NLEs. Above the
step temperature more NLEs are excited. The second fact is that
the NLEs completely disappear at NLE energies of around 1.5 \cite{fw2}.
This corresponds to an energy per particle of 0.5. It is rather
close to the found maximum in the KSE at $T=0.4$.

\section{Discussion}

When the KSE of a system becomes
zero at a certain value of the system control parameter, the system
itself becomes integrable. If the KSE tends to zero (but does not
exactly become) zero approaching a certain range of the
control parameter space, the system becomes close to
an integrable system. Our results show that for the one-dimensional
$\Phi^4$ system at the same time as the KSE(T)-dependence
shows up with a crossover,
the temperature dependence of certain relaxation times of the system
does the same.
What that result implies is that if the relaxation of a system
drastically slows down, the system itself becomes drastically close
to an integrable system. That means that the system evolves over longer
and longer times close to some tori in the phase space, mixing occurs
on larger time scales. The mixing time scale which should be essentially
identical with the relaxation time becomes separated from the
time scale provided by the motion on the tori of the corresponding
integrable system (i.e. the inverse frequencies in the action-angle
representation of the integrable system).

Now we can formulate an essential part of our results. If
the relaxation time of a system becomes drastically large
(by smooth changes of control parameters) that would imply
that certain Lyapunov coefficients may tend to zero. But surprisingly
we find that the largest Lyapunov coefficient tends to zero,
and thus the whole KSE. Consequently the whole system becomes close
to an integrable system.

In analyzing the data for $C=4$ we found that in the
temperature region of low KSE and large relaxation times
the system becomes close to a weakly interacting kink-phonon
system. Thus our analysis provides us with an understanding
of the typical 'quasiparticles' (phonons, kinks) and
the reasons for slow relaxation (low kink density, weak
phonon-phonon interaction). The $C=0.1$ case turns out to
be similar in the correlation between KSE and relaxation, but
totally different in the interpretation of the excitation
spectrum. Here our analysis supports the picture of
nonlinear localized excitations as 'quasiparticles' together
with kinks. The slow relaxation is now given by the slow diffusion
of kinks (high Peierls-Nabarro barrier) and the slow relaxation/interaction
of NLEs.

Butera and Caravati \cite{bc87} have found numerically that a system
of the Heisenberg O(2) universality class shows up with a
crossover of the maximum Lyapunov coefficient versus temperature
behaviour at the phase
transition (where both correlation length and relaxation times
diverge). This result could be viewed in analogy to our
$C=4$ case. Undoubtly
the system becomes nonergodic if one passes the critical
temperature from above. The strange part in both
results is: the fact that the largest Lyapunov coefficient
(and thus the KSE) tend to zero at the critical point
implies the system to be close
to an integrable one!

Let us also mention the results of Pettini and Landolfi
\cite{pl90} and Pettini \cite{mp93}. These authors have investigated
a modified $\Phi^4$ model (where $V(X)=\frac{1}{2}X^2+\frac{1}{4}X^4$)
and Fermi-Pasta-Ulam models in one dimension. All these models
seem not to have a phase transition, thus the found slowing down
can not be attributed to large spatial correlations. The observed
crossover both in the relaxation times and in the largest
Lyapunov exponent versus temperature dependence
were thus attributed to the presence of
a strong stochasticity threshold. This threshold separates
motion mainly along resonances from motion mainly across resonances
of an assumed underlying and perturbed integrable system.

Another interesting case are the studies of liquid-glass transitions. These
transitions are defined by a slowing down of structural relaxation
in the undercooled liquid. Recently Madan and Keyes \cite{mk93} have studied
the
dynamics of Lennard-Jones liquids. Using molecular dynamics they
calculated the fraction of unstable modes out of an averaged
density of states. Around the freezing (glass) transition
a crossover in the temperature dependence of the fraction of unstable modes
is found. Although there is no clear mathematical connection
between their density of states and the Lyapunov spectrum, it seems to
be likely that an investigation of the largest Lyapunov exponent
would yield analogous results.

Summarizing we have shown two examples of slowing down in complex
systems. The simultaneous decrease of the KSE allows us to
make statements about the nature of excitations in the systems
under consideration. Below the crossover (dynamical phase
transition, strong stochasticity threshold) the systems are likely to
behave as a set of weakly interacting excitations. Thus one
can construct microscopic theories to describe the
crossover phenomena.
\\
\\
\\
Acknowledgements
\\
\\
It is a pleasure to thank C. R. Willis and U. Bahr for
stimulating and helpful discussions. We thank the computational
center of the Technische Universit\"at Dresden for support.
This work was supported in part (SF) by the Deutsche Forschungsgemeinschaft
(Fl 200/1-1).

\newpage

\begin{tabbing}
FIGURE CAPTIONS
\normalsize
\\
\\
\\
FIG.1  \hspace{20pt} \= Position of the low-frequency relaxational peak \\
\> of $\chi_{X_lX_l}^{''}$ $\omega_{\alpha '}$ (cf. \cite{fs93}) versus \\
\> temperature $T$ for $C=4$, $N=4000$. \\
\\
\\
\\
FIG.2 \> Kolmogorov-Sinai entropy $S_{KS}$ versus temperature $T$ \\
\> for $C=4$; \\
\> triangles - $N=20$, squares - $N=50$, circles - $N=100$. \\
\\
\\
\\
FIG.3 \> $T^{0.5}/\xi $ versus $1/T$ for $C=4$, $N=4000$.\\
\\
\\
\\
FIG.4 \> $T^{0.5}S_{KS}$ versus $1/T$ for $C=4$. \\
\> Symbols same as in Fig.2. \\
\\
\\
\\
FIG.5 \> Same as in Fig.1 but for $C=0.1$. \\
\\
\\
\\
FIG.6  \> Same as in Fig.2 but for $C=0.1$. \\
\> Insert: Zoom of the temperature region $0 < T < 0.15$. \\
\\
\\
\\
FIG.7 \> Local normalized energy-energy correlator $S_{e_le_l}$ \\
\> versus time for $C=0.1$, $N=1000$ and the temperatures \\
\> $T=0.055/0.067/0.07/0.073/0.11/0.15$.  \\
\\
\\
\\
FIG.8 \> Hypsometric plot for the displacement patterns for $C=0.1$ \\
\> and $N=1000$. A segment of the chain (100 particles) is actually \\
\> shown. Ordinate - time, Abscissa - particle number. A filled square \\
\> is drawn if the given particle has negative displacement. \\
\> Else white space is left. The thin white stripes between neighboring \\
\> square sequences are of no importance.\\
\\
\\
\\
FIG.9 \> Same as in Fig.8 but: a filled square is drawn if \\
\> the displacement of the given particle is closer to any of the \\
\> groundstate positions $\pm 1$ than 0.3.\\
\\
\\
\\
FIG.10 \> $\chi_{e_le_l}^{''}$ for $C=0.1$, $N=1000$ versus \\
\> frequency $\omega$ as calculated from the correlators in Fig.7. \\
\end{tabbing}

\end{document}